\documentclass [a4paper,12pt]{article}
\usepackage{amssymb}
\usepackage{graphicx}
\usepackage{amsmath}
\usepackage{fancybox}
\usepackage{color}
\usepackage{geometry}

\newcommand{\be}{\begin{equation}}
\newcommand{\ee}{\end{equation}}
\newcommand{\bea}{\begin{eqnarray}}
\newcommand{\eea}{\end{eqnarray}} 
\begin{document}
\setlength{\baselineskip}{18pt}
\begin{titlepage}

\begin{flushright}
KOBE-TH-18-02     
\end{flushright}
\vspace{1.0cm}
\begin{center}
{\LARGE\bf Majorana neutrino masses in the scenario of gauge-Higgs unification} 
\end{center}
\vspace{25mm}

\begin{center}
{\large
K. Hasegawa 
and C. S. Lim$^*$
}
\end{center}
\vspace{1cm}
\centerline{{\it
Department of Physics, Kobe University,
Kobe 657-8501, Japan.}}

\centerline{{\it
$^*$
Department of Mathematics, Tokyo Woman's Christian University, Tokyo 167-8585, Japan }}
%
%

\vspace{2cm}
\centerline{\large\bf Abstract}
\vspace{0.5cm}

In this paper we consider possible mechanisms to generate small Majorana neutrino
masses for active neutrinos in the scenario of gauge-Higgs unification, 
a candidate for physics beyond the standard model. We stress that it is non-trivial 
to find a gauge-invariant operator, responsible for the Majorana masses, which is 
the counterpart of the well-known SU(2)$_L \times$ U(1)$_Y$ invariant 
higher-mass-dimensional ($d = 5$) operator. 
As the first possibility we discuss the seesaw 
mechanism by assigning leptonic fields to the adjoint representation of 
the gauge group, 
so that a $d = 5$ gauge-invariant operator can be formed. It turns out that the 
mechanism leading to the small Majorana  masses is the admixture of the Type I and 
Type III seesaw mechanisms. As the second possibility, we consider the case where 
the relevant operator has $d = 7$, by introducing a matter scalar belonging to the 
fundamental representation of the gauge group. Reflecting the fact that the mass 
dimension of the operator is higher than usually expected, the Majorana masses 
are generated by a ``double seesaw mechanism."

\end{titlepage} 

\section{Introduction} 

The standard model (SM) possesses a few serious theoretical problems. A well-known important 
problem is 
that of gauge hierarchy. The attempts to solve this problem have led to representative 
scenarios of physics beyond the standard model (BSM). The most well-studied scenario is 
supersymmetry, whose concrete realization is minimal supersymmetric standard model (MSSM).  
In this paper we focus on the scenario of gauge-Higgs unification (GHU), where the Higgs 
boson is originally a gauge boson. To be precise, the Higgs field is identified with  
(the Kaluza-Klein (KK) zero mode of) an extra-dimensional component of a higher dimensional 
gauge field \cite{Manton:1979kb}, \cite{Hosotani}. A nice feature of this scenario is 
that, by virtue of the higher-dimensional local gauge symmetry, the quantum correction to 
the Higgs mass is UV-finite, thus opening a new avenue for the solution of the hierarchy problem
\cite{Hatanaka:1998yp}. 

Another basic theoretical problem in the standard model is that there is no principle to 
restrict Higgs interactions, such as Yukawa couplings. Namely, there is no guiding principle 
to determine the quark and lepton masses theoretically. From this viewpoint, again the GHU 
scenario is hopeful: the GHU scenario may provide a natural mechanism to restrict Higgs 
interactions, relying on the gauge principle. Let us note that in GHU, Yukawa couplings are 
originally gauge coupling since the Higgs field is originally a gauge field. 

Neutrino masses are expected to play special roles in the investigation of the viability of 
the various BSM scenarios. First, it should be noticed that if neutrinos are assumed to be 
Majorana fermions, the neutrino mass matrix (in the basis of weak eigenstates) is directly 
determined by the observed neutrino mass eigenvalues, generation mixing angles, and (physical) 
CP phases, some of them having already been fixed (with some errors) or restricted 
experimentally. Thus it is possible to compare the prediction of each BSM scenario with 
such determined mass matrix. This is in contrast to the case of quark mass matrices; 
here, because of the freedom of unitary transformations in the sector of right-handed 
quarks, even though we know all of the observables mentioned above, the mass matrices 
cannot be uniquely fixed. 

It should also be noticed that the mass matrices of the lepton sector show very 
characteristic features: neutrino mass eigenvalues are remarkably small compared to those of 
quarks and charged leptons. Also impressive is that (two of the) generation mixing angles in 
the Maki-Nakagawa-Sakata matrix are considerably greater than the corresponding angles in the 
Kobayashi-Maskawa matrix. These interesting features may also have their origin in the 
Majorana nature of neutrinos. Let us recall that only neutrinos, being electrically neutral, 
can be Majorana fermions without contradicting charge conservation.   

Based on these observations as the first step, in this paper we study  
systematically how small Majorana neutrino masses can be realized in the GHU scenario. 
In the literature, the most popular mechanism for realizing small neutrino masses is
the seesaw mechanism \cite{Seesaw}. In this paper, we will propose possible models to 
realize this idea concretely in the framework of GHU. 

What is special for the GHU scenario in the discussion of Majorana neutrino masses ? 
We may easily note that to realize the aforementioned mechanisms for inducing small Majorana 
neutrino masses in GHU is a little challenging. First, since the Higgs field is 
originally a gauge field belonging to the adjoint representation of the gauge group, 
it is non-trivial to form a gauge-invariant operator with mass dimension $d = 5$ 
(from a four-dimensional (4D) point of view) corresponding to the well-discussed 
$\mbox{SU(2)}_{L} \times \mbox{U(1)}_{Y}$ invariant operator, $(\phi^{\dagger}L)^{2}$, where $L$ 
is a left-handed lepton doublet and $\phi$ is the Higgs doublet. Also, the Yukawa coupling 
coming from the covariant derivative of higher-dimensional gauge theory usually preserves 
fermion number, and to break the lepton number is a non-trivial task, though if we extend 
our discussion to the grand GHU \cite{GrandGHU}, the gauge interactions there may lead 
to the violation of baryon and/or lepton number.

\section{Seesaw mechanism in the GHU scenario} 

We discuss how the seesaw mechanism is realized in the GHU scenario, by taking the minimal 
unified electro-weak GHU model, i.e. the 5D SU(3) model \cite{Kubo}. The extra dimension is 
assumed to be an orbifold $S^{1}/Z_{2}$ in order to break SU(3) into the 
gauge group of the SM and also to realize a chiral theory. Let us note that in GHU the 
gauge group of the SM should inevitably be enlarged, and the simplest choice is SU(3). 
The Higgs field behaves as an octet, the adjoint representation of SU(3). Then, in this model, 
assigning leptonic fields in an SU(3) triplet will be unrealistic. First, the charge 
assignment in this model is such that the fields in the triplet all have fractional 
charges, being identified with those of quark fields. 
(The situation will change if the gauge group has an additional U(1) factor 
\cite{Adachi:2014wva}.)
Second, the $d = 5$ operator $(A_{y}\,L)^{2}$  ($L$: lepton 
triplet, $A_{y}$: the fifth component of the 5D gauge field) contained in 
$(D_{y}L)^{2}$ ($D_{y}$ denotes the gauge covariant derivative), which should be the 
counterpart of $(\phi^{\dagger}L)^{2}$, clearly cannot be gauge invariant.

For these reasons, we assign lepton fields to an SU(3) octet $\Psi$, whose 
component fields have integer charges. Also, by taking this choice of representation, we can 
immediately find a $d = 5$ operator ${\rm Tr}\{ [A_{y}, \Psi] ^{2} \}$ stemming from 
a gauge-invariant operator ${\rm Tr}\{ (D_{y} \Psi)^{2} \}$ (with spinor indices being 
omitted for brevity), responsible for the Majorana mass of $\nu_{L}$. One may wonder 
why we do not introduce an SU(3) singlet field to be identified with $\nu_{R}$. Let us 
note that such a $\nu_{R}$ cannot form a Dirac mass with $\nu_{L}$ through the 
vacuum expectation value (VEV) of $A_{y}$, since $\nu_{R}$ and $\nu_{L}$ belong to 
different representations of gauge group and therefore cannot communicate with each another 
through Yukawa coupling. The octet $\Psi$ possesses 
both of the SU(2)$_L$ doublet containing $\nu_{L}$ and the SU(2)$_L$ singlet containing 
$\nu_{R}$ in a single representation.     

In order to complete the seesaw mechanism, in addition to the Dirac mass term mentioned 
above, the Majorana mass term for $\nu_{R}$ is needed. At first glance the Majorana 
mass term seems to be provided by a gauge-invariant operator ${\rm Tr} \Psi^{2}$. 
Unfortunately, the story is not so straightforward. First, although the adjoint representation 
is a real representation and therefore it seems to be natural to assign Majorana particles to 
this representation, it is known that we cannot regard each component of $\Psi$ as an ordinary 
4D Majorana field, satisfying 
the Majorana condition 
$\psi^{c} = \psi$ with $\psi^{c} = C (\bar{\psi})^{t} \ (C = i\gamma^{0}\gamma^{2})$ 
being 4D charge conjugation, even though the number of components of the spinor is the same 
as in the case of 4D space-time. It may be worth noting the fact that there does 
not exist a Majorana spinor in 5D space-time. 
In fact, even if we try to form a Majorana mass term for a generic four-component spinor 
$\psi$, $\overline{\psi^{c}}\psi$, the mass term is known to be non-invariant under the 
5D Lorentz transformation, which  connects 4D space-time coordinates with the extra 
space coordinate.  

Interestingly, we realize that if we add $\gamma_{5}$ to the mass term to form 
$\overline{\psi^{c}}\gamma_{5}\psi$, the modified mass term turns out to be invariant 
under the full 5D Lorentz transformations. So, the linear combination 
$\psi + \gamma_{5} \psi^{c}$ seems to be a self-conjugate spinor, correctly transforming 
under the 5D Lorentz transformation. Unfortunately this is not the case, since 
$\gamma_{5} (\gamma_{5} \psi^{c})^{c} = - \psi$. However, this in turn means that once 
we form an eight-component spinor $\psi_{SM}$,
\be 
\label{2.0} 
\psi_{SM} = 
 \left(
    \begin{array}{cc}
      \psi \\
      \gamma_{5} \psi^{c}  
    \end{array}
  \right), 
\ee 
it is self-conjugate in the following sense: 
\be 
\label{2.1} 
\psi_{SM} 
  =  \left(
    \begin{array}{cc}
      0 & -1 \\
      1 & 0 
    \end{array}
  \right) 
  \gamma_{5}
   (\psi_{SM})^{c}.     
\ee 
$\psi_{SM}$ represents for a ``symplectic Majonara" spinor \cite{Kugo}. Just as the 5D SUSY 
gauge theory can be naturally obtained from the SUSY (pure) Yang-Mills theory in 6D space-time 
by naive dimensional reduction, it may be useful to construct a Lagrangian for $\psi_{SM}$, 
as if the space-time is 6D, and then perform a naive dimensional reduction to 5D. Adopting 
the following basis for 6D gamma matrices,  
\be 
\label{2.2}
\Gamma^{\mu} = \gamma^{\mu} \otimes \sigma_{1}, \ \Gamma^{5} = i \gamma_{5} \otimes \sigma_{1}, \ 
\Gamma^{6} = -i I_{4} \otimes \sigma_{2} 
\ee 
(with the 6D chiral operator being given by $\Gamma_{7} = I_{4} \otimes \sigma_{3}$), 
the symplectic Majorana condition reads 
\be
\label{2.3} 
\psi_{SM} = \Gamma^{5} \Gamma^{6} \Gamma^{2} (\psi_{SM})^{\ast}.   
\ee

In 5D space-time with the $S^{1}/Z_{2}$ orbifold as the extra dimension, the $Z_{2}$ transformation 
is a sort of chiral transformation from the 4D point of view, and hence $\psi$ and $\psi^{c}$ 
should have opposite $Z_{2}$ parities. Thus, the 4D Majorana spinor is not compatible with the 
orbifolding. (This is a reflection of the fact that in 4D space-time there is no 
Majorana-Weyl spinor.) For the symplectic Majorana spinor, the $Z_{2}$ 
transformation can be modified into that in the orbifold $T^{2}/Z_{2}$, the extra dimension 
of 6D space-time:   
\be 
\label{2.4} 
Z_{2}: \ \ \psi_{SM}(x^{\mu}, y) \ \ \to \ \ P^{-1} (-i)\Gamma^{5}\Gamma^{6}\psi_{SM}(x^{\mu}, -y) P   
\ee 
where $y$ is the extra-dimensional coordinate and the $3 \times 3$ matrix $P$ defines 
the $Z_{2}$-parities of each component of the fundamental representation of SU(3), i.e. triplet, as  
\be 
\label{2.5} 
P = {\rm diag} (1, 1, -1). 
\ee 
Since $-i \Gamma^{5}\Gamma^{6} = \gamma_{5} \otimes \sigma_{3}$, $\psi$ and $\psi^{c}$ are 
now allowed to have opposite 4D chiralities, as they should. The transformation 
$-i \Gamma^{5}\Gamma^{6}$ is a rotation of an angle $\pi$ in the 2D extra dimension. 
So, it should be equivalent to ordinary $Z_{2}$ transformation after the dmensional 
reduction to the 5D space-time. In fact, it is easy to check that the bilinear form 
$\overline{\psi_{SM}} \Gamma^{M} \psi_{SM}$ for $M = \mu \ (\mu = 0\,\mbox{--}3), \ 5$, the 
transformation in Eq.(\ref{2.4}) is equivalent to the transformation without $\Gamma^{6}$.   

If the 4D spinor $\psi$ of Eq.(\ref{2.0}) only has the left-handed Weyl spinor $\psi_{L}$, 
for instance as the result of the orbifolding mentioned above, 
\be 
\label{2.6} 
\psi_{SM} = 
\left(
    \begin{array}{cc}
      \psi_{L} \\ \hline 
      (\psi_{L})^{c} \\  
    \end{array}
  \right) 
= 
\left(
    \begin{array}{cc}
      0 \\ \hline 
     \bar{\eta}_{\dot{\alpha}} \\ \hline
      \eta^{\alpha} \\ \hline 
      0 \\  
    \end{array}
  \right) \ \ (\alpha, \ \dot{\alpha} = 1, 2),  
\ee 
which just reduces to the four-component 4D Majorana spinor $\psi_{M}$: 
\be 
\label{2.6'} 
\psi_{M} = 
\psi_{L} + (\psi_{L})^{c} 
= 
 \left(
    \begin{array}{cc}
      \eta^{\alpha} \\ \hline
      \bar{\eta}_{\dot{\alpha}} \\   
    \end{array}
  \right). 
\ee
Then the mass term for $\psi_{SM}$ just reduces to the 4D Majorana mass term for $\psi_{M}$: 
\be 
\label{2.7}
M \overline{\psi_{SM}} \psi_{SM} = M \overline{\psi_{M}} \psi_{M},   
\ee 
where $\overline{\psi_{SM}} = \psi_{SM}^{\dagger} \Gamma^{0}$, while 
$\overline{\psi_{M}} = \psi_{M}^{\dagger}\gamma^{0}$.      

Now we are ready to discuss our model more concretely. The SU(3) octet $\Psi$ contains 
symplectic Majorana spinors $\psi_{SM}^{a} \ (a = 1\mbox{--}\,8)$ as its component fields:  
\be 
\label{2.8} 
\Psi = \psi_{SM}^{a} \frac{\lambda_{a}}{2}, \ \ 
\psi_{SM}^{a} = 
\left(
    \begin{array}{cc}
      \psi^{a} \\
      \gamma_{5}(\psi^{a})^{c}  
    \end{array}
  \right),
\ee
where $\lambda_{a} \ (a = 1\mbox{--}\,8)$ are Gell-Mann matrices. The free Lagrangian for $\Psi$ 
with Majorana mass $M$, after naive dimensional reduction into 5D space-time, is given as 
\be 
\label{2.9} 
{\cal L}_{{\rm free}} = {\rm Tr} \left\{ \bar{\Psi} 
\left(\,\sum_{M = 0\,\mbox{--}3, 5}i\partial_{M} \Gamma^{M} - M \right) \Psi \right\}  
\ee 
Let us note that the condition in Eq.(\ref{2.3}) is compatible with 5D Lorentz transformation, 
but not compatible with the Lorentz transformation connecting a sixth (extra space) coordinate 
with 5D coordinates, reflecting the fact that in 6D space-time there does not exist 
a Majorana spinor. So Eq.(\ref{2.9}) is invariant only under 5D Lorentz transformation, 
which is sufficient for our purpose. The gauge interaction of $\Psi$ is described by 
\be 
\label{2.10} 
{\cal L}_{{\rm int.}} = 2g {\rm Tr} 
\left\{ \bar{\Psi} \sum_{M = 0\,\mbox{--}3, 5}A_{M} \Gamma^{M} 
\frac{1 + \Gamma^{7}}{2} \Psi \right\}. 
\ee 

As the result of the orbifolding, the sector of the KK zero mode is given as follows 
(we show only the part with +1 eigenvalue of $\Gamma_{7}$, $\Psi^{(+)}$):
\bea  
\Psi^{(+)} &=& \Psi^{(+)}_{L} + \Psi^{(+)}_{R}, \nonumber \\ 
\Psi^{(+)}_{L} &=& 
\frac{1}{\sqrt{2}}
 \left(
 \begin{array}{c|c|c}
 0  & 0   & \tilde{e}^{+}   \\ \hline
 0  & 0   & \tilde{\nu}   \\ \hline
 e^{-}  & \nu  & 0 \\
 \end{array}
 \right)_{L}, \ \ 
 \Psi^{(+)}_{R} = 
 \left(
 \begin{array}{c|c|c}
 \frac{N_{\gamma}}{\sqrt{3}}  & \frac{\tilde{E}^{+}}{\sqrt{2}}  & 0   \\ \hline
 \frac{E^{-}}{\sqrt{2}}  & -\frac{N_{\gamma}}{2\sqrt{3}} - \frac{N_{Z}}{2}  & 0   \\ \hline
 0  & 0   &  -\frac{N_{\gamma}}{2\sqrt{3}} + \frac{N_{Z}}{2}   \\
 \end{array}
 \right)_{R},   
 \label{2.11}  
\eea 
where $N_{\gamma}, \ N_{Z}$ are associated with the generators, which are identical to 
those for the neutral gauge bosons $\gamma, \ Z$, and hence both are mixtures 
f the SU(2) singlet (associated with $\lambda_{8}$), corresponding to $\nu_{R}$, and 
triplet (associated with $\lambda_{3}$) leptons. 

Also relevant is the KK zero mode of the extra-dimensional component of the gauge field, 
$A_{y}$: 
\begin{align} 
\label{2.11'}
 A_{y} =
 \frac{1}{\sqrt{2}}
 \left(
 \begin{array}{c|c|c}
 0  & 0  & \phi^{+}   \\ \hline
 0  & 0  & \phi^{0}   \\ \hline
 \phi^{-} & \phi^{0\ast} &  0  \\
 \end{array}
 \right)\,,
\end{align}
where $(\phi^{+}, \phi^{0})$ is nothing but the Higgs doublet in the SM, whose VEV, 
$\langle \phi^{0} \rangle = \frac{v}{\sqrt{2}}$, spontaneously breaks the gauge symmetry of 
the SM through the Hosotani mechanism \cite{Hosotani}.  

Now let us move to the discussion of how the seesaw mechanism is realized in this model. 
For that purpose, we restrict our discussion to electrically neutral leptons. 
Our task is to realize the small Majorana mass for $\nu_{L}$, belonging to the SU(2) doublet 
$L = (\nu , e^{-})_{L}$, through the seesaw mechanism. For the mechanism to work, 
the ``exotic" left-handed doublet $\tilde{L} = (\tilde{\nu}, \tilde{e}^{+})_{L}$ is 
redundant, as it does not exist in the standard model. Also, if it remains in the low 
energy effective theory, it will form a gauge-invariant Dirac mass term with the 
doublet $(\nu, e^{-})$ of our interest,  
\be 
\label{2.12} 
M \left\{ 
\left(
 \begin{array}{cc}
 \overline{\nu_{L}}  & \overline{e^{-}_{L}} \\
 \end{array}
 \right) 
 \left(
 \begin{array}{c}
 (\tilde{\nu}_{L})^{c} \\ 
 (\tilde{e}^{+}_{L})^{c} \\ 
 \end{array}
 \right) + \ h.c. \right\}. 
\ee
The coefficient $M$, supposed to be the mass scale of the $\nu_{R}$ Majorana mass, is assumed to 
be much larger than the weak scale $M_{W}$ for the seesaw mechanism to work, 
and hence $\nu_{L}$ will decouple from our low-energy world. A possible way out of this 
problem is to introduce a brane-localized SU(2) doublet $L_{b}= (\nu_{b}, \ e_{b}^{+})_{R}$ 
to form a brane-localized Dirac mass term at one of the fixed points of the orbifold 
(where the gauge symmetry is reduced to that of the SM by the orbifolding),   
\be 
\label{2.13} 
M_{b} \left\{ 
\left(
 \begin{array}{cc}
 \bar{\nu}_{bR}  & \bar{e^{+}}_{bR} \\
 \end{array}
 \right) 
 \left(
 \begin{array}{c}
 \tilde{\nu}_{L} \\ 
 \tilde{e}^{+}_{L} \\ 
 \end{array}
 \right) + h.c. \right\}. 
\ee 
The ``brane-localized mass" $M_{b}$ is assumed to be much larger than the Majorana mass 
$M$, $M_{b} \gg M$. In an extreme limit, $M_{b}  \to  \infty$, $\tilde{\nu}$ is completely 
decoupled from the theory forming a Dirac mass with $\nu_{b}$, thus leaving $\nu$ alone as 
a massless state. By the way, all the fields appearing in Eqs.(\ref{2.11})--(\ref{2.13}) 
should be understood to be 4D fields with proper mass dimension and kinetic terms after 
the dimensional reduction to 4D space-time. 

After the spontaneous gauge symmetry breaking due to 
$\langle \phi^{0} \rangle = \frac{v}{\sqrt{2}}$, $\nu_{L}$ belonging to the SU(2) doublet 
$L$ forms a Dirac mass term of order $gv \sim M_{W}$ with a right-handed neutral lepton 
in Eq.(\ref{2.11}), behaving as either a singlet or a triplet of SU(2)$_L$ ($2 \times 2 = 1 + 3$). 
It turns out that the partner of $\nu_{L}$ to form the Dirac mass is $N_{Z}$ (not $N_{\gamma}$). 
This is basically because the VEV of $A_{y}$ is electrically neutral and hence 
$[Q, \langle A_{y} \rangle] = 0$, with 
$Q = {\rm diag}(\frac{2}{3}, \ -\frac{1}{3}, -\frac{1}{3})$ being the charge operator 
associated with $N_{\gamma}$. $N_{\gamma R}$, being isolated from other states, obtains a 
Majorana mass $M$ by itself from the Majorana mass term 
$-M {\rm Tr} (\bar{\Psi} \Psi)$ in Eq.(\ref{2.9}).   

From these lessons, we learn that though we have five neutral leptonic states to start with, 
$\nu_{L}, \  \tilde{\nu}_{L}, \ \nu_{bR}, \ N_{\gamma R}, \ N_{ZR}$, in the limit 
$M_{b} \ \to \ \infty$, the Majorana mass of $\nu_{L}$ is effectively determined by 
the diagonalization of the mass matrix $M_{2\times 2}$ in the basis of subsystem 
$(\nu_{L}, \ (N_{ZR})^{c})$: 
\be 
\label{2.14} 
{\cal L}_{m} = - \frac{1}{2}
\left(
 \begin{array}{cc}
 \overline{(\nu_{L})^{c}} & \overline{N_{ZR}} \\
 \end{array}
 \right) M_{2\times 2} 
\left(
 \begin{array}{c}
 \nu_{L} \\ 
 (N_{ZR})^{c} \\ 
 \end{array}
 \right), \ \ \ 
M_{2\times 2} = \left(
    \begin{array}{cc}
      0 & \sqrt{2} M_{W} \\
      \sqrt{2} M_{W} & M 
    \end{array}
  \right). 
\ee 
The mass eigenvalues (their absolute values) are well approximated under $M_{W} \ll M$ to be 
$M$ and $\frac{2M_{W}^{2}}{M}$. The smaller mass $\frac{2M_{W}^{2}}{M}$ is the Majorana mass 
for the mass eigenstate, which is nearly $\nu_{L}$.   

This completes the seesaw mechanism. Though not shown here, we have investigated the diagonalization 
of the full $5 \times 5$ mass matrix and have confirmed that under the condition 
$M_{W} \ll M \ll M_{b}$ we obtain the approximate result, identical to the one mentioned above.   

The physical reason for getting the small Majorana mass for $\nu_{L}$ is the decoupling of 
$N_{Z}$ due to its large Majorana mass. An important remark here is that the state $N_{Z}$ 
is a mixture of SU(2) singlet and triplet states. Namely, the seesaw mechanism operating 
in this model is the admixture of two types of seesaw mechanism, i.e. Type I \cite{Seesaw} 
and Type III \cite{Foot:1988aq}.

\section{Majorana neutrino masses due to \\ higher-mass-dimensional operator} 

As was discussed in the introduction, in GHU it seems to be impossible to form a 
gauge-invariant operator corresponding to the dimension $d = 5$ and 
$\mbox{SU(2)}_{L} \times \mbox{U(1)}_{Y}$ 
invariant operator, $(\phi^{\dagger}L)^{2}$, when the lepton doublet $L$ is assigned as 
a member of the fundamental representation of the gauge group. Actually, this is based on our 
implicit assumption that only the extra-dimensional component of the gauge field $A_{y}$ 
is the field developing the VEV, which breaks the gauge symmetry of the standard model. 
Once we relax this constraint and allow the introduction of a scalar field, which also develops 
a VEV, the situation will change. If the VEV of the introduced scalar is a singlet concerning 
the gauge group of the SM, it has nothing to do with the weak scale $M_{W}$ and hence the 
gauge hierarchy problem. By introducing such a scalar field, together with the lepton 
multiplet and $A_y$, it will become possible to form a gauge-invariant operator with mass 
dimension $d$ higher than 5, typically\,7.   

As an example to make this idea concrete, we discuss a 5D SU(4) unified electro-weak GHU model. 
In this model, the Higgs doublet is contained in the triplet representation of the sub-group SU(3), 
in contrast to the case of the SU(3) model in the previous section, where the Higgs doublet is a 
member of the SU(3) octet. As a result, in this model the predicted weak mixing angle is a 
desirable one, $\sin^{2} \theta_{W} = 1/4$ \cite{Hasegawa}.  The fields responsible for the 
Majorana neutrino mass generation are denoted as follows: 
\be 
\label{3.1} 
A_{y} = 
\left(
    \begin{array}{ccc|c}
      -\frac{2}{\sqrt{6}}a_{Z} & w^{+} & w^{++} & \phi^{0} \\
      w^{-} & -\frac{1}{\sqrt{2}}a_{\gamma} + \frac{1}{\sqrt{6}}a_{Z}  & 
\tilde{w}^{+} & \phi^{-} \\ 
      w^{--} & \tilde{w}^{-} & \frac{1}{\sqrt{2}}a_{\gamma} + \frac{1}{\sqrt{6}}a_{Z}  
& s^{+} \\ \hline 
      \phi^{0 \ast} & \phi^{+} & s^{-} & a_{Z'} \\ 
    \end{array}
  \right), \ \ 
\psi = 
 \left(
    \begin{array}{c}
      \nu_{L} \\
      e^{-}_{L} \\ 
      e^{+}_{L} \\ \hline 
      \nu_{R} \\ 
    \end{array}
  \right), \ \    
\Phi = 
 \left(
    \begin{array}{c}
      \hat{\phi}^{0} \\
      \hat{\phi}^{-} \\ 
      \hat{\phi}^{+} \\ \hline 
      \hat{s}^{0} \\ 
    \end{array}
  \right), \ \ 
\chi^{0}.   
\ee 
We have introduced a scalar field $\Phi$. Leptons have now been assigned to the fundamental 
representation of SU(4), whose chirality is tentative and will be fixed by the orbifolding discussed 
below. We also introduce a gauge singlet fermion $\chi^{0}$. 

The Lagrangian, relevant for the neutrino mass, is 
\be 
\label{3.2} 
{\cal L} = g \bar{\psi}A_{y} \psi + \epsilon (y) M_{b} \bar{\psi} \psi +  
\alpha (\bar{\psi} \Phi \chi^{0} + h.c.) + \frac{1}{2}M \{ \bar{\chi^{0}} \gamma_{5} 
(\chi^{0})^{c} + h.c. \},  
\ee 
where $\epsilon (y)$, with $y$ being the coordinate of the extra space assumed to be an 
orbifold $S^{1}/Z_{2}$, is the ``sign function": $\epsilon (y) = 1\,  {\rm and} -1$ for 
positive and negative $y$, respectively, and the ``$Z_2$-odd" bulk mass term 
$\epsilon (y) M_{b} \bar{\psi} \psi$ causes exponential suppression of the Yukawa 
coupling $f \simeq g (\pi RM_{b}) e^{-\pi RM_{b}}$ ($R$: the radius of $S^{1}$), 
which is desirable in order to get a small Dirac mass, relevant for the lighter 
(first or second) generation.   

The breaking SU(4) $\to$ SU(3) $\to$ SU(2)$_{L} \times$ U(1)$_Y$ due to the orbifolding 
is realized by adopting the following assignment of the $Z_{2}$ parities at two fixed 
points of the orbifold $S^{1}/Z_{2}$ for the fundamental representation of SU(4): 
$P = {\rm diag} (++, ++, +-, --)$. 

The remaining KK zero modes as the result of the orbifolding are 
\be 
\label{3.3} 
A_{y} = 
\left(
    \begin{array}{ccc|c}
      0 & 0 & 0 & \phi^{0} \\
      0 & 0 & 0 & \phi^{-} \\ 
      0 & 0 & 0 & 0 \\ \hline 
      \phi^{0 \ast} & \phi^{+} & 0 & 0 \\ 
    \end{array}
  \right), \ \ 
\psi = 
 \left(
    \begin{array}{c}
      \nu_{L} \\
      e^{-}_{L} \\ 
      0 \\ \hline 
      \nu_{R} \\ 
    \end{array}
  \right), \ \    
\Phi = 
 \left(
    \begin{array}{c}
      0 \\
      0 \\ 
      0 \\ \hline 
      \hat{s}^{0} \\ 
    \end{array}
  \right), \ \ 
\chi^{0}_{L}.   
\ee 
We have put the overall phase as $-1$ for $\Phi$ in its $Z_{2}$ transformation, so that only 
$\hat{s}^{0}$ has the even parity $(+, +)$. 

The higher-mass-dimensional ($d = 7$) gauge-invariant operator relevant for the Majorana 
mass of $\nu_{L}$ is known to be 
\be 
\label{3.4}
\left( \bar{\psi} A_{y} \Phi \right)^{2}. 
\ee
It is easy to confirm that this operator does contribute to the Majorana mass after 
$A_{y}$ and $\Phi $ are replaced by their VEV: $\langle \phi^{0} \rangle = v, \ 
\langle \hat{s}^{0}  \rangle= V$. 

Here some comments are in order on the issue of how the scalar field $\Phi$ can develop 
its non-zero VEV, $V$. One possibility is just to add a gauge-invariant potential term 
for $\Phi$, $- \mu^{2} \Phi^{\dagger}\Phi + \lambda (\Phi^{\dagger}\Phi)^{2} \ (\mu^{2}, 
\ \lambda > 0)$. We notice, however, that this may cause a problem, since the 
mass-squared term for $\Phi$ is not protected by any symmetry and the radiatively induced 
VEV, $V$, may be quite large (unless we perform some fine-tuning). This in turn may 
cause too-large spontaneous symmetry breaking of SU(4) (though the breaking of SU(4) 
itself is realized by orbifolding as was discussed above), leading to a too-large 
positive mass-squared for the Higgs doublet, belonging to the ``broken generator" 
of SU(4), through the term $(D_{M}\Phi)^{\dagger} (D^{M} \Phi)$. Thus, realizing 
the spontaneous breaking of the SM gauge symmetry and getting a small Higgs mass 
becomes non-trivial. 

Hence, a desirable alternative choice would be to embed the $\Phi$ field as part 
of $A_{y}$ by adopting a larger gauge group, so that the potential of $\Phi$ becomes 
under control thanks to the higher-dimensional gauge symmetry. Two VEVs, $V$ and $v$, 
both being generated by the Hosotani mechanism \cite{Hosotani}, may be naturally comparable 
in their orders of magnitude. 

After the spontaneous breaking, the mass terms relevant for $\nu_{L}$ can be read off 
from the Lagrangian in Eq.(\ref{3.2}), and are summarized here in a form using a mass matrix: 
\be 
\label{3.5} 
- \frac{1}{2} 
\left(
 \begin{array}{ccc}
 \overline{(\nu_{L})^{c}}  & \overline{\nu_{R}} & \overline{\left( \chi^{0}_{L} \right)^{c}} \\ 
 \end{array}
 \right) 
\left(
    \begin{array}{ccc}
      0 & fv & 0  \\
      fv & 0 & \alpha V \\ 
      0 & \alpha V & M \\ 
    \end{array}
  \right) \ \ 
  \left( 
    \begin{array}{c}
      \nu_{L} \\
      \left(\nu_{R}\right)^{c} \\ 
      \chi^{0}_{L} \\  
    \end{array}
  \right) + h.c. 
\ee 
We assume a hierarchical structure, $fv \ll \alpha V \ll M$. $fv \ll \alpha V$ is achieved 
by the exponentially suppressed small Yukawa coupling $f$, and the mass scale $M$ can be 
much larger than $v$ and $V$, since it is a singlet with respect to the SM gauge symmetry.  
Then, the diagonalization of the 3$\times$3 mass matrix is straightforward. Namely, 
by an orthogonal rotation in the 2$\times$2 submatrix for the lower two components with the small 
angle of ${\cal O}(\alpha V/M)$, we get the approximate form 
\be 
\label{3.6} 
\left(
    \begin{array}{ccc}
      0 & fv & 0  \\
      fv & - \frac{(\alpha V)^{2}}{M} & 0 \\ 
      0 & 0 & M \\ 
    \end{array}
  \right).  
\ee
We now immediately get three approximated mass eigenvalues (their absolute values), 
$M, \ \frac{(\alpha V)^{2}}{M}$ and 
\be 
\label{3.6'} 
\frac{(fv)^{2}}{\left( \frac{(\alpha V)^{2}}{M}\right)}, 
\ee 
which is identified with the Majorana mass of $\nu_{L}$. In this derivation we have 
made the additional assumption, $(fv)M \ll (\alpha V)^{2}$. Here are two steps of 
a seesaw-like mechanisms, which are seen schematically in Fig.1. Figure\,1 clearly 
shows that the operator giving rise to the Majorana mass is the one given in Eq.(\ref{3.4}). 
A similar mechanism to get the small Majorana neutrino mass has been discussed in 
Ref.\cite{Yamatsu}.   

\begin{figure}[htbp]
 \begin{center}
  \includegraphics[width=100mm]{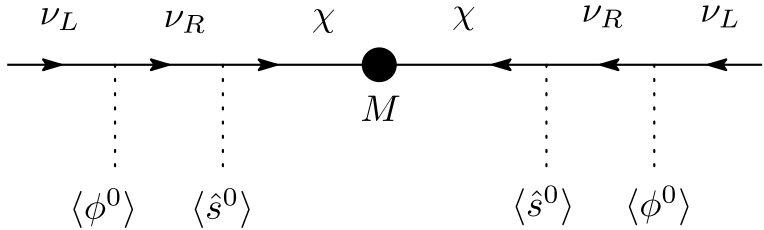}
 \end{center}
\caption{The diagram contributing to Eq.(\ref{3.6'})}
\end{figure}

\section{Summary and discussion} 

In this paper we have considered possible mechanisms to generate small Majorana neutrino 
masses for (active) left-handed neutrinos in the scenario of gauge-Higgs unification, 
one of the attractive scenarios of physics beyond the standard model. A specific feature 
of this scenario in the construction of the Majorana mass term is that it is non-trivial 
to find an operator, responsible for the Majorana neutrino masses, which is a counterpart 
of the SU(2)$_L \times$ U(1)$_Y$ invariant higher-mass-dimensional $d = 5$ 
(from 4D point of view) operator $(\phi^{\dagger}L)^{2}$ ($L$: left-handed lepton doublet, 
$\phi$: Higgs doublet). For instance, in the minimal unified electro-weak SU(3) 
GHU model \cite{Kubo}, we cannot get a gauge-invariant operator by just replacing the 
lepton doublet $L$ by a triplet field, the fundamental representation of SU(3), since the Higgs 
field in this scenario, corresponding to $\phi$, is $A_{y}$ (the extra-dimensional 
component of the gauge field), which of course belongs to the adjoint representation
 of the gauge group. 

As the first possible mechanism to generate small Majorana neutrino masses we discussed 
the seesaw mechanism \cite{Seesaw}. Leptonic matter fields are assigned to the adjoint 
representation of SU(3), i.e. the SU(3) octet, so that the component fields have integer charges and 
a $d = 5$ operator ${\rm Tr}\{ [A_{y}, \Psi] ^{2} \}$ stemming from a gauge-invariant 
operator ${\rm Tr}\{ (D_{y} \Psi)^{2} \}$ can be formed. Though the Majorana spinor seems 
to fit naturally to the octet, i.e. the real representation of the gauge group, it has been known 
that in 5D space-time there does not exist a Majorana spinor. Thus we formulated the 
Lagrangian for leptons by use of the symplectic Majorana spinor \cite{Kugo}, which has 
eight components and naturally fits 6D space-time. Interestingly, in our model the 
partner of $\nu_{L}$ to form a Dirac mass is the admixture of SU(2)$_L$ singlet 
(corresponding to $\nu_{R}$) and triplet fermions. Thus the mechanism operating 
in this model based on the GHU scenario turns out to be the admixture of 
Type I \cite{Seesaw} and Type III \cite{Foot:1988aq} seesaw mechanisms. 

As the second possibility we considered the case where Majorana neutrino masses 
are generated in a form of higher-mass-dimensional ($d > 5$) gauge-invariant operator. 
We argued that once the implicit constraint that the VEV to break the gauge symmetry 
should be given only by the VEV of $A_{y}$ is relaxed, introducing a matter scalar 
belonging to the fundamental representation of the gauge group (together with an additional singlet 
fermionic field), we can form a higher-mass-dimensional ($d = 7$) gauge-invariant 
operator, responsible for the Majorana neutrino masses. Reflecting the fact that the 
relevant operator has a mass dimensional higher than usually expected, the Majorana 
neutrino masses are generated by, say, the double seesaw mechanism.

\subsection*{Acknowledgments}

This work was supported in part by the Japan Society for the Promotion of Science, Grants-in-Aid for Scientific Research, No.~16H00872, No.~15K05062.


\providecommand{\href}[2]{#2}\begingroup\raggedright\endgroup

\end{document}